\documentclass[useAMS,usenatbib]{mn2e}
\usepackage{float,graphics,epsfig,array,amsmath,amssymb}
\usepackage{times}
\usepackage{color}

\newcommand{\flux}{{erg~cm$^{-2}$~s$^{-1}$}}
\newcommand{\ltsima}{$\; \buildrel < \over \sim \;$}
\newcommand{\simlt}{\lower.5ex\hbox{\ltsima}} 
\newcommand{\gtsima}{$\; \buildrel > \over \sim \;$}
\newcommand{\simgt}{\lower.5ex\hbox{\gtsima}} 

\newcommand{\ion}[2]{\ensuremath{\mbox{#1~{\sc #2}}}}
\newcommand{\errUD}[2]{\ensuremath{^{+#1}_{-#2}}}

\newcommand{\chidof}{\ensuremath{\chi^2/\mbox{d.o.f.}}}

\newcommand{\fekb}{\mbox{Fe  K$\beta$}}

\newcommand{\feka}{\ensuremath{\mbox{Fe~K}\alpha}}

\newcommand{\fexxv}{\ensuremath{\mbox{\ion{Fe}{xxv}}}}
\newcommand{\fexxvi}{\ensuremath{\mbox{\ion{Fe}{xxvi}}}}

\newcommand{\xmm}{{XMM-\emph{Newton}}}

\newcommand{\lum}{erg~s$^{-1}$}
\newcommand{\nh}{cm$^{-2}$}

\newcommand{\chandra}{{\emph{Chandra}}}

\newcommand{\suzaku}{{\emph{Suzaku}}}

\newcommand{\nhsym}{$N_{\rm{H}}$}

\newcommand{\sorg}{Mrk~348}

\def\araa{ARA\&A}%
\def\apj{ApJ}%
\def\apjl{ApJ}%
\def\apjs{ApJS}%
\def\aap{A\&A}%
\def\aaps{A\&AS}%
\def\mnras{MNRAS}%
\def\pasp{PASP}%
\def\pasj{PASJ}%
\def\physrep{Phys.~Rep.}%
\title{The variable ionized absorber in the Seyfert 2 Mrk~348}
\author[E. Marchese et al.]{E. Marchese$^{1}$\thanks{E-mail:elena.marchese@brera.inaf.it}, V.  Braito$^{1}$\thanks{E-mail:valentina.braito@brera.inaf.it}, J. N Reeves$^{2,3}$\thanks{E-mail:j.n.reeves@keele.ac.uk}, R. Della Ceca$^{4}$,
 A. Caccianiga$^{4}$,
 \newauthor  A. Markowitz$^{5,6,7}$, G. Risaliti $^{8,9}$, P. Severgnini$^{4}$, T.J. Turner$^{3}$ \\
$^{1}$INAF - Osservatorio Astronomico di Brera, Via Bianchi 46 I-23807 Merate (LC), Italy \\
$^{2}$Astrophysics Group, School of Physical and Geographical Sciences, Keele University, Keele, Staffordshire, ST5 5BG, UK  \\
$^{3}$Department of Physics, University of Maryland Baltimore County, 1000 Hilltop Circle, Baltimore, MD 21250, USA \\
$^{4}$INAF-Osservatorio Astronomico di Brera, via Brera 28, 20121 Milano, Italy  \\
$^{5}$University of California, San Diego, Center for Astrophysics and Space Sciences, 9500 Gilman Dr., La Jolla, CA 92093-0424, USA \\
$^{6}$ Karl Remeis Sternwarte, Sternwartstrasse 7, 96049 Bamberg, Germany  \\
$^{7}$ Alexander von Humboldt Fellow  \\
$^{8}$ Harvard - Smithsonian Center for Astrophysics, 60 Garden Street, Cambridge, MA 02138, USA  \\
$^{9}$INAF - Osservatorio Astrofisico di Arcetri, Largo E. Fermi 5, 50125 Firenze, Italy }

\begin{document}

\date{}

\pagerange{\pageref{firstpage}--\pageref{lastpage}} \pubyear{2002}

\maketitle

\label{firstpage}

\begin{abstract}

We present the results of the analysis  of the X-ray spectrum of the Seyfert 2 \sorg, observed by \suzaku\ and \xmm.  The overall spectrum of \sorg\ can be described by a primary power law continuum seen through three layers of absorption, of which one is neutral and two are ionised. Comparing \suzaku\ (2008) and \xmm\ (2002) observations we find    variability of the X-ray spectral curvature. We suggest that the variability can be explained through the change of column density of both the neutral and one of the  ionised absorbers, together with a variation of the ionisation level of the same absorber. We thus confirm one of the main features presented in past works, where intrinsic column density variability   up to $\sim 10^{23}$~cm$^{-2}$ was observed on time scales of months. We also find that the photon index of the underlying power law continuum ($\Gamma \sim 1.8$) is  in agreement with the previous observations of this Seyfert 2.
\\

\end{abstract}

\begin{keywords}
galaxies: active -- galaxies: individual (\sorg) --  X-rays: galaxies -- galaxy \sorg 
\end{keywords}

\section{Introduction}
\label{intro}
The extreme energetic phenomena occurring in the nuclei of active galaxies (Active Galactic Nuclei, AGN) are now recognized to be the result  of accretion of large amounts of gas  onto a central supermassive black hole (SMBH), which grows and emits radiation covering a wide range of energies. \\
  In the  Unified Model of AGN \citep{Antonucci} their different observed properties  are explained through orientation effects between our line of
sight to the nucleus and ``circum-nuclear material". This circum-nuclear gas imprints  features -  low energy cut-offs,
the  Compton hump and  emission and absorption lines - onto the primary X-ray
emission.  However, recent studies have shown that AGN are more complex than the simple picture of the Unified Model, where the absorbing matter is uniformly distributed in a toroidal geometry, and located at a pc scale distance from the central region \citep{Antonucci,Urry1995}.  In fact, this  representation does not fully explain the features observed in all AGN \citep{Bianchi2012,Turner09,Turner2012,Elvis2012}, this is the reason why the geometry,
size and physical state of the circum-nuclear medium of AGN are still a matter of
debate. \\
 Indeed, recent X-ray observations of nearby and bright AGN showed the co-existence of multiple absorbing components, tracing gas covering a wide range of column densities (\nhsym) and ionisation states, all of them contributing in giving shape and complexity to the X-ray spectrum we observe (see \citealt{Turner09}). 


The variability observed in the X-ray spectra of   few
nearby AGN  showed that this matter is highly structured with a range of
ionisation states, densities, geometries and locations (\citealt{Turner09},
\citealt{Risaliti2010a}) and, in particular, that a significant fraction of the absorbing medium must be clumpy.  $N_{\rm H}$ variations  have been discovered on various time-scales, allowing to constrain the location and   size of the obscuring clumpy material. There is an increasing number of  obscured or type 2 AGN displaying variability in the X-ray absorbers: NGC~1365 \citep{Risaliti05, Risaliti07, Risa2009a, Maiolino2010}, NGC~4388 \citep{Elvis04}, NGC~7674 \citep{Bianchi05},
NGC~4151 \citep{Puccetti}, NGC~7582 \citep{Xue98,Turner2000,Bianchi09}, UGC~4203 \citep{Risaliti10}, Cen A \citep{Rivers11}, NGC~454 \citep{Marchese2012} and NGC~4507 \citep{Braito2013}. A few of these sources represent the most extremes cases, known  as ``changing-look'' AGN, which show rather strong variability, from a Compton-thin  state (\nhsym =$10^{23}
\rm cm^{-2}$) to a Compton-thick state (\nhsym  $>10^{24} \rm cm^{-2}$),  
on time-scales from a few days down to a few hours.  The short time scale of some of these occultation events, when detectable, implies  that the obscuring clouds are located in the inner region of the AGN, thus in proximity of the accretion disk or in the Broad Line Region. However, the majority of observations show variations on longer times scales (of the order of months or year), due either to observational limitations  or to the intrinsic nature of the source. Moreover, it is not always straightforward to discern if the X-ray variability is due to a change in the column density of the absorber  or to a variation in the primary emission of the X-ray source. Neverthless, it is now known that variability of the X-ray absorbers is a common propriety in type 2 AGN \citep{Risaliti2002} and its analysis can give multiple information on the structure of the circumnuclear matter.   \\\
\sorg , as we will discuss below, does not belong to the class of  changing-look AGN, since  it was not observed in a state with \nhsym$>10^{24} \rm cm^{-2}$; nevertheless  the variability in the absorbing column density along the line of sight allows us to infer that the  absorbing material is not homogeneous and stable, but that has to be clumpy. The data discussed here give us information about the geometry, structure and possible location of the absorbers.
\\

Markarian 348 (NGC 262) is a Compton-thin Seyfert 2 galaxy at $z$=0.015 \citep{Khachikian74,Vaucouleurs}.  The first X-ray observation of this source was performed by \emph{Ginga}, and provided evidence of an absorbed (\nhsym  $\sim 10^{23} \rm cm^{-2} $) X-ray source, with photon index
$\Gamma \sim$1.7 \citep{Warwick89}. In terms of  the unification schemes for Seyfert galaxies, the identification of \sorg\ as a Seyfert 2 was further confirmed in a work published by \cite{Miller1990}  showing that this source is characterised by a broad
(FWHM $\sim$ 7400 km s$^{-1}$) H$\alpha$-line component in polarized light.   \sorg\ is a relatively strong radio emitter and was  observed by the Very Large Array (VLA) and the Multi-Element-Radio-Linked-Interferometer-Network (MERLIN, \citealt{Unger1984,Anton2002}).  It is characterised  by a flat radio spectrum continuing well into the infrared, as well as a core-dominated radio structure and  rapid radio variability (\citealt{Neff1983}). The complex radio properties of this source and the difficulties in determining its orientation with respect to the line of sight make its classification into a radio-quiet or radio loud source uncertain  \citep{Simpson1996}.\\

\sorg\ was  observed with the \emph{Rossi X-Ray Timing Explorer} (\emph{RXTE}) mission in twelve  observations during the period December 29, 1996 to July 12, 1997. \cite{Smith2001} analysed these spectra and described the resulting
time--averaged 3--20 keV spectrum by a power-law continuum ($\Gamma\sim$ 1.8) absorbed by a column density of \nhsym\ $\sim 10^{23} \rm cm^{-2}$,  
plus a \feka\ emission line with equivalent width  $EW\sim$ 100 eV, plus a Compton reflection component. They found  variations in the intrinsic column density  occurring over periods of typically weeks to months, with the
largest change ($\Delta N_{\rm H}  \sim 10^{23}\rm cm^{-2}$), taking place on a time-scale of $\sim$ 70 days. They also found X-ray continuum variations with the shortest observed timescale of $\sim$ 1 day.
 \cite{Smith2001} also found that the \feka\ line flux did not change significantly during the multiple observations, deducing that  much of the line emission is produced in a layer of material with a rather constant sky coverage and thickness, as viewed from the nucleus. They modelled the Compton reflection component using \textsc{pexrav} \citep{pexrav}, finding that the data were consistent with a reflection strength of $R\sim 0.3-0.8$. The 2--10 keV luminosity measured in  these observations was, depending on the  \nhsym\  of each observation, in the range  0.8--3.4$\times 10^{43} \rm erg \ s^{-1}$  (for H$_{0}$ = 50; q$_{0}$ = 0.5), a factor 3 higher than that measured by \emph{Ginga}.\\
 \cite{Smith2001} suggested that the absorber in \sorg\ could consist of individual clouds; motions in and out of the line of sight could explain the observed variations in \nhsym.  \cite{Akylas2002} analyzed the same data as \cite{Smith2001} but with additional 25 \emph{RXTE} observations taken in May--June 1996. This analysis confirmed the spectral variability already observed by \cite{Smith2001}. 
Finally, a more recent work by \cite{Singh2011} on the X-ray spectral properties of a sample of Seyfert galaxies,   analysed the 0.5--10 keV \xmm\ EPIC-pn spectrum of \sorg\ (2002). They suggested the presence of two absorbers intercepting the primary radiation, a  fully covering absorber, with $N_{\rm H}\sim 7 \times 10^{22}$\nh and a partial covering component, with $N_{\rm H}\sim1 \times 10^{23}$\nh\ and covering fraction of $C_{f}\sim 0.84$. They also found a narrow \feka\ line with $EW\sim$34 eV.\\
Here we present the results of a \suzaku\ observation (June 2008, net exposure $\sim$ 76 ks) and  the comparison  with \xmm\ observation (July 2002, net exposure  $\sim$ 31 ks for EPIC-pn). We show that the spectrum of \sorg\  can be described by a primary power-law continuum intercepting  multiple absorbing components, with different ionisation states. We found that the variability of the \nhsym\ of these layers can explain the observed differences between the \xmm\ and \suzaku .\\

The structure of the paper is the following: in Section \ref{data}  we describe \suzaku\ and \xmm\ observations and data  reduction; in Section \ref{suz_spect} we focus on the modelling of the 0.6--70 keV \suzaku\ spectrum to understand the nature of the X-ray absorption and to evaluate the contribution of the reflection and \feka\ emission line on the primary continuum. In Section \ref{xmm_spect} we compare our \suzaku\ best-fit model with the \xmm\ spectrum and analyse the variations in some of the best-fit parameters, i.e. \nhsym\ and the ionisation parameters. Discussion and conclusions follow in Sections \ref{discussion} and \ref{conclusions}. Throughout this paper, a
 cosmology with $H_0$ = 70 km s$^{-1}$ Mpc$^{-1}$ , $\Omega_{\Lambda}$=0.73,
and $\Omega_{\rm m}$=0.27 is adopted.

\section{Observations and data reduction}
\label{data}
\subsection{\suzaku}
\label{suzaku}
 \sorg\ was observed by  the Japanese X-ray satellite \suzaku\ (\citealp{Mitsuda07}) on  28th June 2008     for a
total exposure time of about 88 ks.\\
\suzaku\ carries on board four  X-ray Imaging Spectrometers (XIS, \citealp{Koyama07}), with X-ray CCDs at their focal plane, and a non-imaging hard X-ray detector (HXD-PIN, \citealp{Takahashi07}). At the time of this observation only three of the XIS were working: one  back-illuminated (BI) CCD (XIS1) and two front-illuminated (FI) CCDs (XIS0 and XIS3).
All together the XIS and the HXD-PIN cover the
0.5--10 keV and 12--70 keV bands respectively. The spatial resolution of the XIS is $\sim$ 2 arcmin (HEW), while the 
field of view (FOV) of the HXD-PIN is   34  arcmin radius.
Data from the XIS and HXD-PIN  were  processed using v2.1.6.14
of the \suzaku\ pipeline  and applying the standard screening parameters\footnote{The screening
filters all  events  within the South Atlantic Anomaly (SAA)  as well as  with an
Earth elevation angle (ELV) $ < 5^{\circ }$ and  Earth day-time
elevation angles (DYE\_ELV) less than $ 20 ^{\circ }$. Furthermore
also data within  256s of the SAA were excluded from the XIS and within 500s of the
SAA for the HXD. Cut-off rigidity (COR) criteria of $ > 8 \,\mathrm{GV}$ for the
HXD data and $ > 6 \,\mathrm{GV}$ for the XIS were used.}.


\subsubsection{The \suzaku\ XIS analysis}

The XIS data were selected in $3 \times 3$ and $5\times 5$ editmodes using only
good events with grades 0, 2, 3, 4, 6 and filtering the  hot and flickering pixels with
the script \textit{sisclean}.  The XIS response (rmfs) and ancillary response (arfs) files were
produced,   using the latest calibration files available, with the \textit{ftools}
tasks \textit{xisrmfgen} and \textit{xissimarfgen} respectively. The net exposure times are 76 ks for each of the XIS.  The XIS  source spectra  were extracted from a circular region of 2.9$'$
 centered on the source,  and the  background spectra  were extracted from two
circular regions with radius 2.3$'$,  offset from the source and the calibration
sources.    The spectra from
the  two FI  CDDs (XIS 0 and XIS 3) were combined to create  a single source
spectrum (hereafter XIS--FI),  while the  BI (the XIS1) spectrum  was kept separate
and fitted simultaneously.   
The net 0.5--10 keV  count rates  are: $(0.793\pm
0.003)$ counts s$^{-1}$, $(0.829\pm 0.003)$ counts s$^{-1}$, $(0.709\pm 0.003)$ counts s$^{-1}$ for the  XIS0,
XIS3 and XIS1  respectively. We considered data in the range 0.6--10 keV for the XIS--FI  and in the range 0.6--9 keV  for the XIS--BI (ignoring the band 1.6--1.9 keV, due to the presence of instrumental calibration
uncertainties). The difference on
the upper boundary for the XIS1 spectra is because this CCD is optimised for the  soft X-ray band, with higher background at higher energies.
The net XIS source spectra were binned at the \suzaku\ energy resolution and then grouped  to a  minimum 
of 20 counts per bin in order to use  $\chi^2$ statistics.   

  \subsubsection{The \suzaku\ HXD-PIN analysis}
For the HXD-PIN data  reduction and analysis we followed the latest \suzaku\ data
reduction guide (the ABC guide Version
2\footnote{http://heasarc.gsfc.nasa.gov/docs/suzaku/analysis/abc/}),   and used the
rev2 data, which include all 4 cluster units.   The HXD-PIN instrument team
provides the background (known as  the ``tuned'' background) event file, which
accounts for  the instrumental  ``Non X-ray Background'' (NXB; \citealt{kokubun}).  The
systematic uncertainty of   this ``tuned'' background model is  
$\pm$1.3\% (at the 1$\sigma$ level for a net 20 ks
exposure, \citealt{Fukuzawa2009}).\\   We extracted the
source and background spectra   using the same common good time interval, and 
corrected the source spectrum  for the detector dead time. The net exposure time 
after    screening  was 73 ks. We  then simulated a spectrum for   cosmic
X-ray background counts \citep{Boldt,Gruber} and added  it to the  instrumental
one. \sorg\ was detected at a level of 36.8$ \%$ above the background. \\
Fitting this spectrum with a power law we obtain a photon index $\Gamma=1.50^{+0.09}_{-0.09}$ and  $F_{(14-70~\mathrm{keV})}\sim 1.5 \times 10^{-10} \rm erg \ cm^{-2} \ s^{-1}$. The extrapolation of this flux in the 14--195 keV band gives F$_{(14-195~\mathrm{keV})}\sim 3.4 \times 10^{-10} \rm erg \ cm^{-2} \ s^{-1}$. This  flux is higher than the \emph{Swift}--BAT flux, which is $F_{(14-195~\mathrm{keV})}\sim 1.6 \times 10^{-10} \rm erg \ cm^{-2} \ s^{-1}$. We also note  that  \emph{Swift}--BAT observation provides a different photon index, $\Gamma\sim 1.9$, probably due to the wider energy range. If we fit simultaneously \suzaku\ HXD-PIN and \emph{Swift}--BAT spectra, fixing the photon indexes to the same value, we find $\Gamma\sim 1.8$;    if we extrapolate the  flux in the 14--195 keV band for \suzaku\ HXD observation we find $F_{14-195 \rm keV}^{HXD}\sim 2.3 \times 10^{-10} \rm erg \ cm^{-2} \ s^{-1}$, while the   flux measured during \emph{Swift}--BAT observations is $F_{(14-195~\mathrm{keV})}^{BAT}\sim 1.6 \times 10^{-10} \rm erg \ cm^{-2} \ s^{-1}$. The variation could be explained by the fact that \emph{Swift}--BAT spectrum is an average of the observations made during 58 months, while HXD observation provides a snapshot of the spectrum, when the flux was possibly higher,  therefore it is not surprising to find intrinsic variations between them.

 \subsection{\xmm}
 
 \xmm\ observed \sorg\ on July 18, 2002  with a total duration of about 50 ks.
The \xmm\ observatory (Jansen et al. 2001) carries, among its onboard
 instruments, three 1500 cm$^2$ X-ray telescopes, each with EPIC (European Photon Imaging Cameras) imaging
spectrometers at the focus. 
Two of the EPIC use MOS CCDs
(\citealt{Turner01})  and one uses a pn CCD (\citealt{Struder01}). These CCDs allow observations in the range
$\sim$0.5--10 keV. 

 During this observation the pn, MOS1, and MOS2 cameras had
the medium filter applied and they were operating in Full Frame Window mode.
The  data have been processed using the Science Analysis Software
(SAS ver. 6.5) and analysed using standard software packages (FTOOLS ver. 6.1). Event files have been filtered for high-background time
intervals, and only events corresponding to patterns 0--12 (MOS1, MOS2) and
to patterns 0--4 (pn) have been used. The net exposure
times at the source position after data cleaning are $\sim$31.5 ks (pn),
$\sim$38.3 ks (MOS1) and $\sim$38.2 ks (MOS2).   The net count rate in the 0.5--10 keV band  is 2.078$\pm$0.008 counts s$^{-1}$ (pn), 0.647 $ \pm$ 0.004 counts s$^{-1}$ (MOS1) and 0.650$\pm$0.004  counts s$^{-1}$ (MOS2). The effect of pile-up for this source is negligible. \\
The results of the analysis of the \xmm\ observation were already published  in past works (\citealt{Singh2011,Guainazzi2011}), describing a best-fit model composed of a fully covering  absorber ($N_{\rm H}\sim 7 \times 10^{22}$\nh), and a partial covering component with  $N_{\rm H}\sim10.5 \times 10^{22}$\nh and covering fraction of $C_{f} \sim 0.84$. Therefore for this work we will only use the pn data (the MOS spectra are consistent with it), in order to compare it to the \suzaku\ spectra.

\section{Spectral analysis}
\label{suz_spect}
All the models were fitted to the data using  standard software packages (\textsc{xspec} ver.
12.6.0, \citealt{xspecref}) and including   Galactic absorption   ($N_{\rm{H,Gal}}=5.86\times 10^{20}$\nh ;
\citealt{Kalberla2005}).    In the following, unless otherwise stated, fit parameters are quoted in the rest
frame of the source at $z$=0.015 and errors are at the 90\% confidence level   for one interesting
parameter ($\Delta\chi^2=2.71$).

\subsection{Suzaku spectral analysis}
\label{suzaku_analysis}
For the analysis we fitted simultaneously   the \suzaku\ spectra  from  the XIS-FI (0.6--10 keV), the 
XIS1(0.6--9 keV) and HXD-PIN (14--70 keV).  We set the cross-normalization factor  between the HXD and the XIS-FI spectra    to 1.16 (allowing it to vary by $\pm 5$\%), as recommended for XIS nominal
observation processed after 2008 July\footnote{http://www.astro.isas.jaxa.jp/suzaku/doc/suzakumemo/suzakumemo-2007-11.pdf;\\
http://www.astro.isas.jaxa.jp/suzaku/doc/suzakumemo/suzakumemo-2008-06.pdf}. The cross normalization factor between the XIS-FI and the XIS1 spectra was allowed to vary.\\ 

\begin{figure}
\begin{center}
 \resizebox{0.5\textwidth}{!}{
\rotatebox{-90}{
\includegraphics{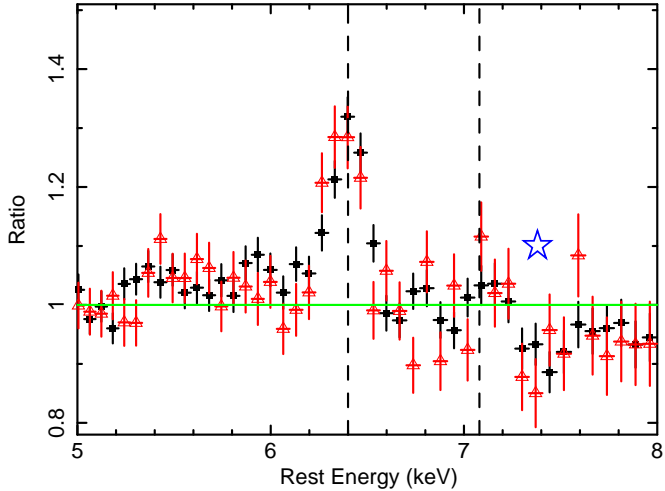}
}}
\caption{Data/model ratio between  the \suzaku\ data (XIS-FI: black filled circles; XIS1: red  open triangles, colours in the electronic version only) and a basic continuum model composed of an absorbed and an additional unabsorbed (scattered) power law,  showing the iron line profile. The
two vertical dashed lines correspond to the rest-frame energies of the \feka \ 
and \fekb\ emission lines at 6.4 keV and 7.06 keV respectively. The star indicates a possible absorption feature at $\sim$ 7.4 keV.}
 \label{res_model2}
\end{center}
\end{figure}

As a starting point  the data were fitted by a simple model composed of an absorbed primary power-law component and an unabsorbed power law, representing the fraction of primary X-ray radiation that is scattered into our line of sight. The photon indices of these two components were tied to each other.  At this first stage we fitted only the continuum, excluding the data between 5 and 7.5 keV where we expected the \feka\ emission complex. This model provided a poor fit with a  $\chi^2$ of 411.4 for 266 degrees of freedom (\emph{d.o.f.}), with a photon index $\Gamma\sim$1.57 and \nhsym $\sim$ 1.24 $\times 10^{23} \rm cm^{-2}$.   When   including   the data in the 5--7.5 keV energy range we obtained   $\chi^2/ \emph{d.o.f.}=$ 950/344. In Figure \ref{res_model2} we report the residuals of this simple model, including all data, which clearly reveal the presence of emission lines at energies corresponding to  the \feka\ ($E\sim 6.4 $ keV),  and \fekb\ ($E\sim 7.06$ keV) emission lines. \\
We thus added two Gaussian components, in order to reproduce these emission lines, fixing the energy of the \fekb\ line to 7.06 keV and its normalization    to be 13.5\% of the \feka\, consistent with the theoretical value \citep{Palmeri}.  The addition of the \feka\ and \fekb\ lines improved significantly the fit, yelding as energy centroid  for the \feka\ line E=$6.39^{+0.01}_{-0.01}$ keV and for the \fekb\ line E=$7.10^{+0.06}_{-0.06}$ keV  and a $\Delta \chi^2=338.4$ for 3 d.o.f, with respect to the first model applied to all the data (including the 5--7.5 keV energy range). This was still not a good fit ($\chi^2 /\emph{d.o.f.}=$ 611.6/341). The equivalent width  of the \feka\ with respect to the observed continuum is $EW=81.5^{+9.0}_{-8.8}$ eV, its energy centroid is $E=6.39^{+0.01}_{-0.01}$ keV and $\sigma=67.9^{+19.1}_{-21.0}$ eV.
 Considering that the Compton thin matter has an \nhsym\ of $\sim 1.2 \times 10^{23}$\nh\, we expect it to produce in transmission an equivalent width of $EW \sim 30$ eV for Solar abundances (\citealt{Murphy2009}). Therefore we can suppose that the measured \feka\ equivalent width is mainly due to   reflection by Compton thick matter located out of the line of sight.   \\
Given the presence of a flat continuum  ($\Gamma\sim 1.57$) and of  the \feka\ emission with an equivalent width of $EW\sim 81$ eV  we included a Compton reflection component. This component was modelled by adding  the \textsc{pexrav} model \citep{pexrav}. Fitting the model including this component we obtained $\chi^2/\emph{d.o.f.}$=594.0/340 (thus $\Delta \chi^2=17.6$ compared to the model with the \feka\ and \fekb\ emission lines). The parameters characterizing the reflected component are: an inclination angle \emph{i} fixed to 60$^{\circ}$, abundance Z=Z$_{\odot}$,  a reflection
fraction (defined by the subtending solid angle of the reflector $R=\Omega/4\pi$) 
found to be $R\sim 0.41$  and a normalisation fixed to the normalisation of the absorbed power law. The fact that the reflection is quite weak is in agreement with the modest equivalent width ($EW\sim 81$ eV)   found for the \feka\ emission line (\citealt{Ghisellini94,Matt96}).\\

\begin{figure}
\begin{center}
 \resizebox{0.5\textwidth}{!}{
\rotatebox{-90}{
\includegraphics{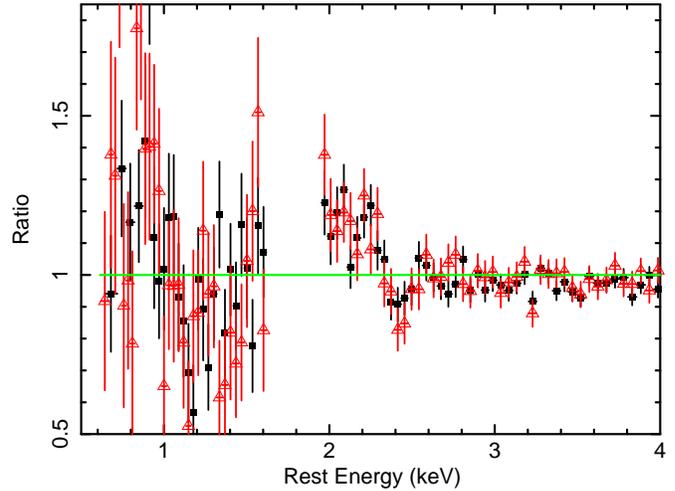}
}}
\caption{Data/model ratio in the soft X-ray energy region to  the \suzaku\ data (XIS-FI: black filled circles; XIS1: red  open triangles, colours in the electronic version only) and   the model including the reflection component (modelled with \textsc{pexrav}). These residuals highlight the possible presence of an emission line at $\sim 2.2$ keV and of an absorption line at $\sim$2.4 keV.  
\label{res_model5b}
}
\end{center}
\end{figure}

%
%
 The model still did not provide a good fit, in particular  between 2 and 2.4 keV, as   can be observed in Figure \ref{res_model5b}. Thus we added  an emission line at about 2.2 keV and an absorption line at about 2.4 keV, even if we cannot exclude that these features could be  instrumental features, such as the Au M edge. This provided a better representation of the soft X-ray spectrum, yielding a  $\chi^2$ of 534.1 for 336 d.o.f. ($\Delta \chi^2$ of 59.9 for 4 d.o.f.). The energy centroids found for the lines are  $E=2.22^{+0.02}_{-0.02}$ keV and $E=2.42^{-0.03}_{+0.03}$ keV. The model achieved until this point gives the following  best-fit parameters: $\Gamma=1.75^{+0.06}_{-0.05} $,  $R=0.41^{+0.17}_{-0.16}$, \nhsym $=1.29 ^{+0.28}_{-0.31}\times 10^{23} \rm cm^{-2}$.  Since there are still some residuals in the soft X-ray band   we also added a thermal component, modelled through  \textsc{mekal} (\citealt{Mewe85})  in \textsc{xspec}.  \\

\begin{figure}
\noindent
 \includegraphics[height=1.0\hsize,angle=270]{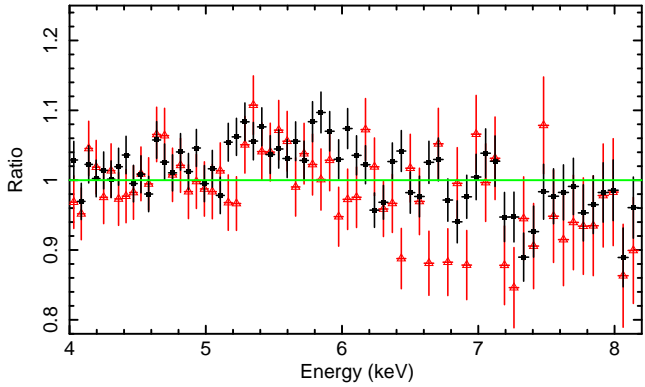}
 \caption{Data/model ratio in the 4--8 keV energy region between  the \suzaku\ data (XIS-FI: black filled circles; XIS1: red  open triangles, colours in the electronic version only) and   the model including the  reflection component (modelled with \textsc{pexmon}).   }
 \label{residui_confronto}
\end{figure}

 We then replaced the neutral reflection component \textsc{pexrav} with   the \textsc{pexmon} model (\citealt{Nandra2007}) in \textsc{xspec} , which self-consistently includes the \feka, \fekb, Ni K$\alpha$ and the \feka\ Compton shoulder.
The parameters characterizing \textsc{pexmon} are: an inclination angle \emph{i} fixed to 60$^{\circ}$, a cut off energy of $ E=$200 keV (\citealt{Dadina08}), a scaling reflection factor $R $ left free to vary, abundance $Z=Z_{\odot}$,   and a reflection normalisation fixed to the normalisation of the absorbed power law. The fit including \textsc{mekal} and \textsc{pexmon} gives a $\chi^2/\emph{d.o.f.}$=514.3/337,  and $R=0.43^{+0.04}_{-0.04}$.  The fit is still poor due to the presence of residuals in the energy range 5--8 keV (see Fig. \ref{residui_confronto}).\\

\subsubsection{Fe K complex}
\label{fek_complex}
In this Section we  discuss  the possible presence of emission lines belonging to the Fe K complex and their properties. We looked for signatures of ionised emission lines, by adding a Gaussian component centered at $E=6.7$ keV and then at $E=6.96$ keV. The fit did not improve significantly and  the upper limit on the equivalent width of each of these lines is less than 10 eV.\\ 
However, if we observe Figure \ref{residui_confronto}, reporting the residuals of the last model, including the \textsc{pexmon} component,  the presence of a residual weak curvature between 5 and 6 keV and weak absorption lines near 7 keV is noticeable. \\
First we tested for the presence of   broadening of the \feka\ emission line, due to relativistic effects manifesting when the line is produced in the inner regions of the accretion disk.
 Starting from the model including \textsc{pexmon}, we added a Gaussian component initially centered at about $E = 6.4$ keV, allowing its centroid energy to vary without any constraint. The addition of this component   could account for the curvature only through a very broad and not so physical emission line  centered at $E \sim 5.6$ keV and with $\sigma \sim 1	$ keV.  We finally attempted to add a component describing more precisely the line emission from a relativistic accretion disk. Thus we added the \textsc{laor} model in \textsc{xspec} (\citealt{laor91}), representing the emission line profile from around a maximally rotating black hole.  The outer radius $R_{out}$ was fixed at 100 R$_g$, the  inclination was constrained to be 
$>45^{\circ}$ (appropriate for a type 2 AGN) and we allowed as free parameters the emissivity index, the innermost radius $R_{in}$ and the normalization. The fit still fails to reproduce the curvature, unless we allow to vary  the inclination, reaching a best-fit value of 25$^{\circ}$ ($\chi^2/\emph{d.o.f.}$=398.7/333), but this produces more than one inconsistency. In fact it represents a face-on orientation (which is  in disagreement with the classification of this source as a Sy2 under classical Sy1/Sy2 unification schemes and the \feka\ line has an equivalent width of $EW \sim 168$ eV which, without considering the narrow \feka\ emission line, would imply a reflection fraction of  $R\sim1$, in contrast with what we find ($R\sim 0.4$). Given that it appears unlikely to originate from a disk line, we   investigated (Section \ref{ion_abs}) if the curvature could be due to a more complex absorber. \\



\subsubsection{Ionized absorber}
\label{ion_abs}

Since, as discussed in section \ref{fek_complex} and shown in  Figure \ref{residui_confronto}, the presence of a  weak residual curvature and of  weak absorption features is clear,   we   evaluated if a more complex  absorber (i.e. partial covering or ionised) was required by the present data.  \\
Firstly, we included a partial covering absorber in the model, in addition to the fully covering one, resulting in a significant improvement of the fit, yielding $\Delta \chi^{2}=111.9$ for 2 d.o.f.. The column density of this absorber is \nhsym=$1.20^{+5.11}_{-2.65} \times 10^{23} $\nh and the covering fraction is $f_{cov}=0.50^{+0.19}_{-0.14}$. Despite this being a better fit we note that the curvature in the 5--6 keV region is still present. We then used the best-fit values obtained by \citep{Singh2011}, who adopted the same model, fixing the covering fraction to $f_{cov}$=0.84 and finding $N_{\rm H, part.cov.}=9.16^{+0.61}_{-0.67} \times 10^{22}$ \nh and  $N_{\rm H, ful.cov.}=6.79^{+0.33}_{-0.28} \times 10^{22}$  \nh. However the fit get worse by $\Delta \chi^{2}=-11.2$ for 1 d.o.f.\\

We thus  consider if this feature of the spectrum can be better described through an ionised absorber.
We added a model representing a photoionized absorber, which  is made using a   multiplicative grid of absorption model  generated with the {\sc xstar} v 2.1 code \citep{xstar}. This grid describes an ionized absorber
parametrised by its turbolence velocity (here we used 5000 km s$^{-1}$), its \nhsym\ and its ionisation parameter, defined as: 
\begin{equation}
\label{ion_abs_eq}
\xi=\frac{L_{ion}}{nR^2}
 \end{equation} 
 where $L_{\rm ion}$ is the ionising luminosity in $erg/s$ between 1--1000 Rydbergs (13.6
eV to 13.6 keV), \emph{n} is the hydrogen number density in  $\rm cm^{-3}$ and $R$ is the radial distance in $cm$ of the absorber
from the ionising source. The outflow velocity is fixed to zero for simplicity. \\
 We obtained a significant improvement, providing a $\chi^2/\emph{d.o.f.}= 385.6/335$ ($\Delta \chi^2=128.7$  for 2 d.o.f with respect to the model described at the end of section \ref{suzaku_analysis}); the parameters obtained with this fit are \nhsym (neutral absorber) $=5.85^{+0.61}_{-0.66} \times 10^{22}$ \rm cm$^{-2}$, \nhsym (ionized absorber) $=1.49^{+0.10}_{-0.09} \times 10^{23} \rm cm^{-2}$ and a photoionization parameter of log$\xi =1.83^{+0.15}_{-0.14} \rm erg \ cm \ s^{-1} $. The addition of this mildly ionised absorber succeeds in reproducing the  curvature in the 5--6 keV energy range. \\

 Despite being a better description of the data, this model still leaves some residual features in the region 6--7.5 keV, in particular  an absorption feature at $\sim7.4$ keV. \\
As a first step we added an absorption line, modelled by an inverted Gaussian, with its centroid and width left free to vary. This yielded $\Delta \chi^2=7.8$ for 3 \emph{d.o.f.}, with energy of the line $E$=7.40$^{+0.07}_{-0.06}$ keV and $\sigma\sim 0.046$ keV, which implies a FWHM of  $\sim$4400 km/s.  
Despite this weak feature is not highly significant, the  most likely  candidate for it is blue-shifted ($v\sim 0.05 c$) absorption  due to the  1s$\rightarrow$2p transition of H-like Fe (E=6.97 keV), while if we assumed a lower ionization state of Fe (i.e. Fe xxv) the corresponding blue-shift would be higher ($v\sim 0.1 c$).\\
 The last consideration lead us to investigate if the weak absorption feature could be well described by substituting the inverted Gaussian component with a second more ionized absorber. At this step, for simplicity,  we tied the outflow velocity of the more ionised absorber to the velocity of the mildly ionised one, leaving it free to vary. The resulting fit with this additional component  provided $\chi^2/\emph{d.o.f.}=369.3/332$ ($\Delta \chi^2=16.4$  for 3 d.o.f), and gives an acceptable model for the absorption feature at $\sim 7.4$ keV. Then we tested for an improvement of the fit when leaving free to vary the outflow velocity of the highly ionised absorber. We obtained that the fit did not statistically improve ($\chi^2/\emph{d.o.f.}=364.6/331$). The same is true when we fixed the outflow velocity of the more ionised absorber to zero. Thus here-after we make the simple assumption that  the two absorbers have the same outflow velocity. 
 The model adopted is of the form\\
\begin{center}
 $F(E) = \mathrm{wabs} \times
[ \mathrm{mekal}+ (\rm{zphabs}\times \rm{ion \ abs1} \times \rm{ion \ abs2} \times \mathrm{pow1}) + \mathrm{pexmon}+ 
\mathrm{pow2}  +\rm{2.2keV \ emis. line} + \rm{2.4 keV\  abs. line}) ]$.
\end{center}
The main parameters of the best fit model are:\nhsym$_{\rm ion1}=1.50^{+0.09}_{-0.09} \times 10^{23} \rm cm^{-2}$,   \nhsym $_{\rm neutral}=4.14^{+0.41}_{-0.40} \times 10^{22}$ cm $^{-2}$,  \nhsym$_{\rm ion2}$ $= 1.41^{+0.80}_{-0.79} \times 10^{23} \rm cm^{-2}$, log$\xi_1$ $=1.63^{+0.08}_{-0.07}$erg  cm s$^{-1}$, log$\xi_2$ $=3.88^{+0.09}_{-0.28} $erg  cm s$^{-1}$ and $v_{\rm ion1,2}=0.057^{+0.006}_{-0.005}c$. We note that, being the \nhsym\ of the two ionised absorbers very similar, it could be due to a stratification of the same absorber with different ionisation states.
This model for \suzaku\ gives a flux in the 2--10 keV energy range of F$_{2-10 \rm keV} \sim 3.60 \times 10^{-11} \rm erg \ cm^{-2}\ s^{-1}$ and an intrinsic luminosity of $L_{2-10 \rm keV}\sim 3.26 \times 10^{43} \rm  erg \ s^{-1}$.\\
The value of measured outflow velocity (0.057$c$) is in agreement with assuming an absorption feature due to   blueshifted \fexxvi\ absorption line.\\

Physically this model describes radiation that intercepts a neutral absorber and two photoionised absorbers that have an outflow velocity of $\sim0.06$c, attenuating the primary AGN emission and producing blueshifted absorption lines.

\begin{figure}
\begin{center}
 \resizebox{0.5\textwidth}{!}{
\rotatebox{-90}{
\includegraphics{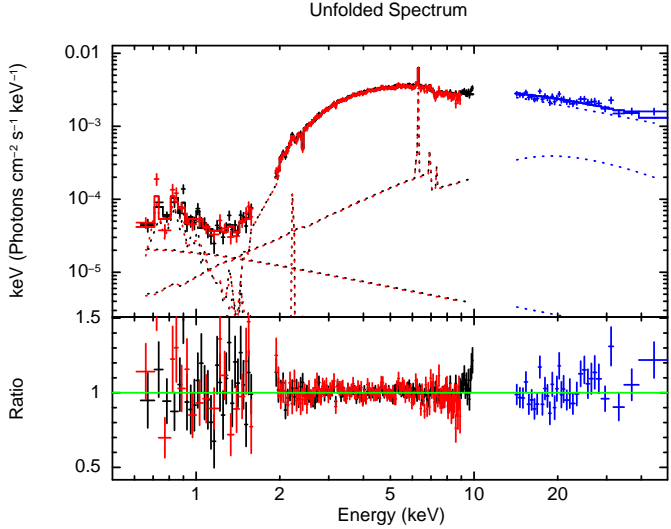}
}}
\caption{\suzaku\ 0.5--70 keV data and best-fit model (in the electronic version: XIS-FI black; XIS1 red; HXD blue) of Mrk 348; data have been rebinned for plotting purposes. The upper panel shows the data and best-fit model described at the end of section \ref{ion_abs}  (see table \ref{tab2} for the best-fit parameters). The lower panel shows the data/model ratio to this model.
\label{euf_suzaku}
}
\end{center}
\end{figure}

\section{\xmm\ data analysis}
\label{xmm_spect}
\sorg\  is candidate for a variable absorber (\citealt{Akylas2002,Smith2001,Singh2011}).
The \nhsym\ measured with \suzaku\ is a factor of 1.8 lower than the \nhsym\ reported by \cite{Singh2011} from \xmm\,  in agreement with the presence of a variable absorber. 
Thus we re-analysed the \xmm\ spectra with some of the models adopted for describing the \suzaku\ spectrum. 

We first considered the  simplest  model, where the reflection is represented by \textsc{pexrav} plus  a Gaussian component for the \feka\ emission line.  When fitting simultaneously the \suzaku\ and \xmm\ data we found that energy centroid of the \feka\ is consistent between the two observations, while its normalization   changed (from $I_{\feka}^{Suzaku}= 4.47^{+0.66}_{-0.65} \times 10^{-5} \rm photons \ cm^{-2} s^{-1} $ and $EW_{\feka}^{Suzaku}=82^{+10}_{-10}$ eV to $ I_{\feka}^{XMM}=2.37^{+0.63}_{-0.61}\times 10^{-5} \rm photons \ cm^{-2} s^{-1}$ and  $EW_{\feka}^{XMM}=47^{+13}_{-12}$ eV). \\

We then adopted the \suzaku\ best-fit model, where the presence of a reflector plus a \feka\ and \fekb\ emission lines was modelled by the \textsc{pexmon} component. Given the observed  variation of the \feka\ emission line intensity we expect that, when applying the best-fit model simultaneously to \suzaku\ and \xmm\ data, we need different reflection scaling factors of \textsc{pexmon}.\\
 In order to test the validity of this hypothesis, we  investigated the following possible scenarios:
\begin{enumerate}
\item  a variable continuum and a constant reflection component, i.e. a scenario in which the reflecting matter is quite far and does not respond immediately to the continuum variability. In terms of fitting parameters this is represented by 
\begin{itemize}
\item a free  normalization of the primary  power law between \xmm\ and \suzaku;
\item the reflection parameters of \xmm\ (reflection fraction $R$ and normalization) tied to those of \suzaku.
\end{itemize}
The resulting fit yielded $\chi^2=567.7$ for 407 d.o.f., and   the assumption of constant reflection leaves residuals in the \feka\ region.
\item  a scenario where both the primary continuum and the reflection component  were allowed to vary between \suzaku\ and \xmm\ observations, i.e. a situation where a variability in the continum reverberates immediately in the reflecting material, which would be located closer in.
Thus the fit was characterised by
\begin{itemize}
\item a free  normalization of the primary power law between \xmm\ and \suzaku;
\item a free reflection fraction $R_{XMM}$ with respect to $R_{\suzaku}$
\item The normalization of \textsc{pexmon} tied to the normalization of the primary power law, for \xmm\ and \suzaku\ independently
\end{itemize}
 This representation lead to a better fit ($\chi^2=551.1$ for 406 d.o.f.),  where the continuum is unchanged (variations within the errors) while the reflection fraction does change, as expected. 
 \end{enumerate}
 
Thus, we conclude that \suzaku\ and \xmm\ observations do not show continuum variability, but the \feka\ emission line does show signs of variation between the two spectra. For this reason we consider hereafter that a good description is obtained by fixing between \suzaku\ and \xmm\ the  continuum and the \textsc{pexmon} normalizations, while allowing the reflection fraction $R$ free to vary between the two observations. \\

\begin{figure}
\begin{center}
 \resizebox{0.5\textwidth}{!}{
\rotatebox{-90}{
\includegraphics{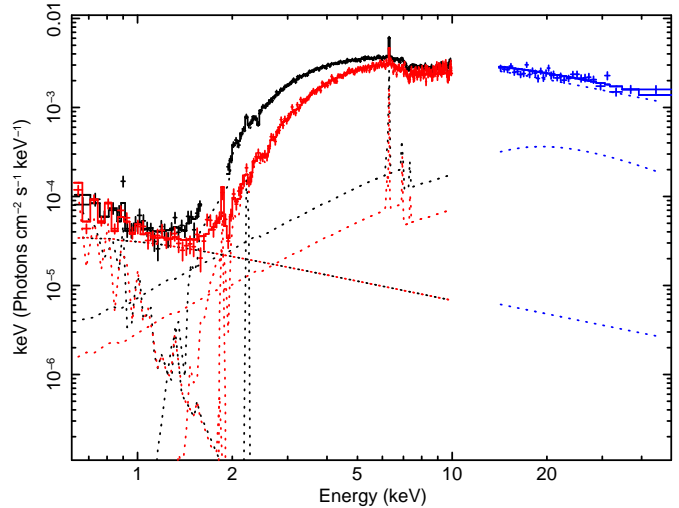}
}}
\caption{
Comparison between the X-ray emission measured with \suzaku\
in 2008 (black data points in the electronic version) and \xmm\ (red
data points in the electronic version) in 2002. The continuum model is the best fit model composed of a primary power-law component transmitted
trough a neutral absorber, a scattered power-law component, a Compton reflected component (modelled with \textsc{pexmon}), and two ionized absorbers.}
\label{confronto_xmm_suzi}
\end{center}
\end{figure}

In Figure \ref{confronto_xmm_suzi} we plot the \suzaku\ XIS (black, upper spectrum), HXD (blue) and \xmm\ pn (red, lower spectrum) data (using  the best-fit model, see below), confirming a change in the spectral  curvature   between 1 and
6 keV. We now analyse what are the spectral components that could be responsible for this variation.
 First of all we tested the  model including the partial covering absorber. Fixing the covering fraction of both \suzaku\ and \xmm\ to $f_{cov}=0.84$ we obtain that  the \nhsym\ changes from  \nhsym$_{part.cov.}^{XMM}=1.55^{+0.12}_{-0.13} \times 10^{23} \rm cm^{-2}$ to \nhsym$_{part.cov.}^{\suzaku}=8.72^{+0.69}_{-0.75} \times 10^{22} \rm cm^{-2}$ for the partial covering absorber, and from  \nhsym$_{ful.cov.}^{XMM}=1.16^{+0.05}_{-0.05} \times 10^{23} \rm cm^{-2}$ to \nhsym$_{ful.cov.}^{\suzaku}=6.93^{+0.36}_{-0.31} \times 10^{22} \rm cm^{-2}$.

Now we consider as a starting point the best-fit model found for \suzaku, thus the model including the ionised absorbers. If we allow as free parameters only the \nhsym\ of the neutral absorber  together with the reflection R factor of  \textsc{pexmon} as well as the soft emission and absorption lines we found $\chi^{2}/\emph{d.o.f.}=589.8/409$. The  neutral absorber column density varies from  \nhsym $_{\rm neutral}^{\suzaku}=4.27^{+0.50}_{-0.47} \times 10^{22} \rm cm^{-2}$ to \nhsym$_{\rm neutral}^{XMM}=1.23^{+0.06}_{-0.06} \times 10^{23}  \rm cm^{-2} $. \\
Since there are some residuals in the 6--8 keV region,  we allowed as free parameters the column densities  of the ionized absorbers.
We obtained $\chi^{2}/\emph{d.o.f.}=568.7/407$ ($\Delta \chi^{2}$=21.1 for 2 d.o.f.); the parameters changing  more significantly are  the \nhsym\ of the neutral absorber the \nhsym\ of the mildly ionised absorber, implying that the source, when observed with \xmm, is generally more absorbed than when observed with \suzaku.\\

We finally allowed also the ionization parameters  to vary between \suzaku\ and \xmm\; we obtained $\chi^{2}/\emph{d.o.f.}=551.6/407$. We found a similar variations for the \nhsym\ of the different absorbers, as those mentioned above.

 The ionisation parameter of the mildly ionised absorber change from log$\xi_{1}^{\suzaku}=1.67^{+0.11}_{-0.10} $erg  cm s$^{-1}$ to log$\xi_{1}^{XMM}=2.04^{+0.19}_{-0.18} $erg  cm s$^{-1}$. This will be considered as the best--fit model and all the relative best-fit parameters  are listed in table \ref{tab2}.

We note that if  we untie the outflow velocity of  \xmm\ ionised absorbers   with respect to \suzaku\ we find $\Delta \chi^2=17$ with respect to the best fit model; the outflow velocity of the first ionised absorber is statistically unchanged, while for  the second ionised absorber it is slightly lower ($z=-0.020^{+0.006}_{-0.006}$ , corresponding to a velocity of 0.034$c$). \\
 We remark that  if we attempt  to fix the \nhsym$_{\rm neutral}$ of \xmm\ to the \suzaku\ best fit value, we get a worse fit by a factor $\Delta \chi^2$=20.   \\

\begin{figure}
\begin{center}
 \resizebox{0.5\textwidth}{!}{
\rotatebox{0}{
\includegraphics{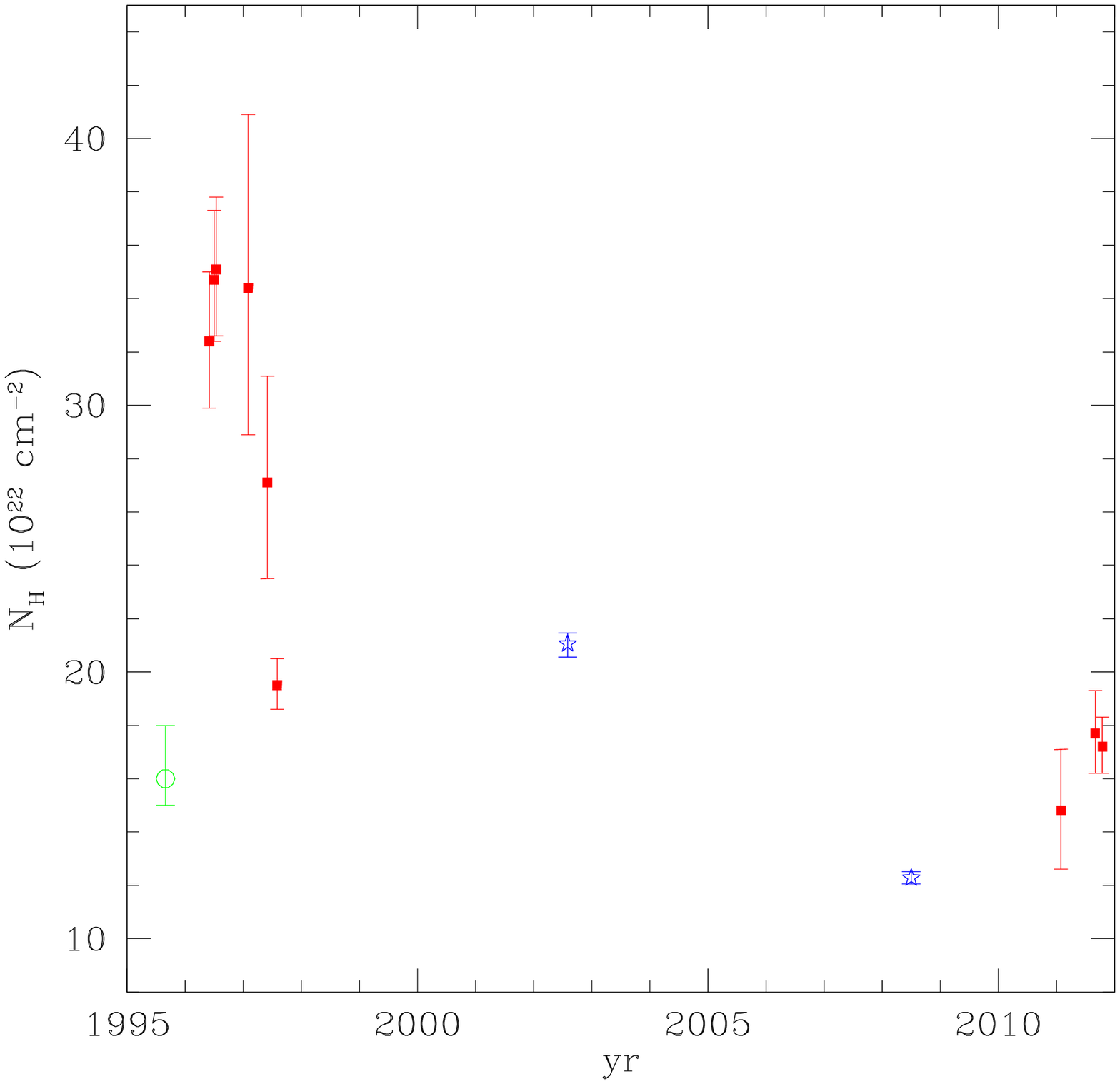}
}}
\caption{Comparison of the measured values of \nhsym in \emph{ASCA} observation (\citealt{Awaki2000},green empty circles), in \emph{RXTE} observation (red squares) during the time lag from 1997 to 2011 (\citealt{Akylas2002} and A. Markowitz in prep. for the 2011 data) and the \nhsym\ measured from \suzaku\ and \xmm\ observations (blue stars, this work). The \nhsym  values of this work and of \emph{RXTE} observations are based on the abundances relative to Solar reported in  \protect \cite{Wilms2000}. } 
\label{confronto_nh}

\end{center}
\end{figure}

We conclude that we can account for the difference in the observed X-ray spectra between \suzaku\ and \xmm\ through  a variation in absorbing column density of both the neutral absorber and one of the ionised absorbers, together with a change in the ionisation parameter of the same ionised absorber. \\
We note that the measured values of \nhsym\  found in this work are within the range of the column densities measured in past observations, in particular: 
\begin{itemize}
\item \emph{Ginga} (1987):  the measured column density was \nhsym $\sim 10^{23}$ \nh
\item \emph{ASCA} (August 1996): as published by \cite{Awaki2000} the column density was  \nhsym\ $\sim 1.6 \times 10^{23}$ \nh. We note that during this observation the observed 2--10 keV flux was $\sim 5 \times 10^{-12} $\flux, that is respectively a factor 5 and a factor 7 lower than what we observed with \xmm\ and \suzaku. If we hypothesize a scenario where there is a delay in \feka\ response with respect to  continum variations, this intrinsic variability in the continuum could be responsible for the different \feka\ intensities discussed at the beginning of Section \ref{xmm_spect}. In fact, the lower intensity \feka\   line observed with \xmm\ could be responding with a time delay to  a different past continum (in this case a weaker continum) in a scenario where the reflector producing this emission line is far from the central variable X-ray source.
\item \emph{RXTE} (from mid 1996 to mid 1997, and in 2011): \nhsym\ took values in the range 0.9 -- 3.2$\times 10^{23}$\nh during 14 months between mid 1996 and mid 1997 (\citealt{Akylas2002,Smith2001}) and in the range 14 -- 17 $\times 10^{22}$ \nh during the observation in 2011 (\citealt{Markowitz2013})

\end{itemize}
We report for clarity in Figure \ref{confronto_nh}  the measured values of \nhsym\ of previous observations together with \xmm\ and \suzaku\ observations.  

 \begin{table*}
\caption{Summary of the X-ray emission lines detected for \suzaku\ spectrum in the 2--8 keV energy range. The energies of the lines are   quoted in the rest
frame. Fluxes and identifications are reported in column  2 and 3.    The $EW$ are reported in column 4 and they are 
calculated against the total  observed continuum at their respective energies. In column 5 the
improvement of fit is shown with respect to the continuum model, the value for the model with no lines 
is \chidof\ =950.0/344.  \label{table:Fe_lines} }
\begin{tabular}{ccccc}
\hline
Energy      & Flux  &ID        & $EW$ & $\Delta \chi^2$  \\
 (keV)       & ($10^{-6}$ph cm$^{-2}$ s$^{-1}$) & &(eV) &\\
 
    (1)   &  (2)   &  (3)    &  (4)   & (5)   \\

\hline
     &     &      &         &  \\
       2.22\errUD {0.02}{0.02}& 8.2\errUD{1.9}{1.9} & S$\rm K \alpha$   & ..     & 54   \\   
 &   &   &     &\\
    2.42\errUD {0.03}{0.03}& -4.6\errUD{2.4}{2.5} & Sxv  & ..    & 22  \\   
 &   &   &     &\\
   6.39\errUD {0.01}{0.01}& 44.8\errUD{5.0}{4.9} & \feka\  & 81.5\errUD {9}{8.8}     & 335    \\   
 &   &   &     &\\

 7.06 & 13.5$\%$ \ \feka  &  \fekb\ & .. & 5\\
  &   &   &      &\\
 
 6.7  & $<$4.8 &  \fexxv &   $<$12 & 0\\
   &   &   &       &\\
 6.97 & $ <$5.8 & \fexxvi  & $<$10  & 0\\
    &   &   &     & \\
 \hline
\end{tabular}
\end{table*}

\begin{table*} 
\caption{Summary of the \suzaku\   and \xmm\  parameters for the best-fit  models described in section
\ref{ion_abs}, and \ref{xmm_spect}. 
\label{tab2}
}

\begin{tabular}{lccc }
\hline
 Model Component  &  Parameter  &  \suzaku &\xmm\ \\ 
  & & 2008-06 & 2002-07 \\
 \hline
&&   \\

Power law &$\Gamma$&$1.72_{-0.02}^{+0.02}$ & fixed to $\Gamma_{\suzaku}$ \\
& Normalisation$^{a}$ & $ 1.60_{+0.08}^{-0.08}  \times 10^{-2} $ & fixed to norm$_{\suzaku}$ \\

Scattered Component  &Normalisation$^{a}$ &$3.66_{-0.42}^{+0.41}\times 10^{-5}$ & fixed to norm$_{\suzaku}$\\
\textsc{mekal}  &Normalisation$^{a}$ &$4.34_{-0.70}^{+0.71}\times 10^{-5}$ & fixed to norm$_{\suzaku}$\\
		  & $k_{\rm B}T$ &$0.24_{-0.02}^{+0.02}$ & fixed to kT$_{\suzaku}$\\
Neutral Absorber   & N$_{\rm H}$& $4.50_{-0.51}^{+0.56}\times 10^{22}$ \nh & $9.99^{-1.39}_{+1.32}\times 10^{22}$ \nh \\

Reflection & R &$ 0.29^{+0.04}_{-0.04} $ & $ 0.12^{+0.05}_{-0.05} $  \\

            	Ionised Absorber 1 & $N_{\rm H}$ & $1.44^{+0.10}_{-0.10} \times 10^{23}$\nh & $2.11^{+0.19}_{-0.18} \times 10^{23}$\nh \\   
               & log$\xi$ & $1.67^{+0.11}_{-0.10} $erg  cm s$^{-1}$ & $2.04^{+0.03 }_{-0.01} $erg  cm s$^{-1}$\\    
               & $z$ & fixed to z of Ionised Absorber 1  &  fixed to z of Ionised Absorber 1  \\
               	& $v_{turb}$& 5000 km $\rm s^{-1}$  & 5000 km $\rm s^{-1}$   \\
               	
   Ionised Absorber 2 &  $N_{\rm H}$ & $1.31_{-0.88}^{+1.22} \times 10^{23}$\nh & $9.94_{-0.57}^{+29.6} \times 10^{22}$\nh \\   
               & log$\xi$ & $3.87^{+0.11}_{-0.39} $erg  cm s$^{-1}$& $ 3.73^{+0.52}_{-0.27} $erg  cm s$^{-1}$\\   
               & $z$ & -0.044$^{+0.007}_{-0.006}$ &   fixed to $z_{\suzaku}$  \\   
               	& $v_{\rm turb}$&  5000 km $\rm s^{-1}$& 5000 km $\rm s^{-1}$   \\
 
&F $_{(0.5-2)\mathrm {keV}}$ &$\sim 3.3\times 10^{-13}$\flux &$\sim  3.4 \times 10^{-13 }$\flux\\
&F$_{(2-10)\mathrm {keV}}$  &$\sim 3.6 \times 10^{-11}$\flux &$\sim 4.2  \times 10^{-11 }$\flux\\
&F $_{(14-150)\mathrm {keV}}$ &$\sim 7.6 \times 10^{-12}$\flux &$\sim 7.04   \times 10^{-12 }$\flux\\
&L $_{(0.5-2)\mathrm {keV}}$  &$\sim 1.72 \times 10^{43}$\lum&$\sim 1.73   \times 10^{ 43 }$\lum \\
&L$_{(2-10)\mathrm {keV}}$  &$\sim 3.26\times 10^{43}$\lum &$\sim 3.17   \times 10^{43}$\lum\\
&L$_{(14-150)\mathrm {keV}}$  &$\sim 4.51 \times 10^{42}$\lum &$\sim 4.20  \times 10^{ 42}$\lum \\
 \hline
\end{tabular}\\
\begin{flushleft}
 
$^a$ units of  photons  keV$^{-1}$ cm$^{-2}$  s$^{-1}$.\\ 

\end{flushleft}
\end{table*}

\section{Discussion}
\label{discussion}
The presence
of one or more ionized absorbers is not  exceptional, indeed recent sensitive
observations with \chandra, \xmm, and \suzaku\  unveiled the presence of  red- and blue-shifted
photoionized absorption lines both in  type 1 and type 2 AGN as well as in Radio Quiet and Radio Loud AGN 
\citep{Tombesi2010b,Tombesi2011,Tombesi2013,Gofford2013}.
 Thus,  it appears that
there is a   substantial amount of  ionized gas in the nuclei of AGNs, which may be
linked  to  gas outflowing on parsec scales with velocities from hundreds of km/s 
up to  $v_{\rm out}\sim 0.04 -0.15c$ \citep{Tombesi2010b,Tombesi12}.\\
It is interesting to make a first order estimate of the maximum distance of this ionised absorber from the central black hole by means of the equation
\begin{equation}
R_{ion}=\frac{L_{\rm ion} \Delta R}{N_{\rm H} \xi R}
\label{ion_abs_eq}
\end{equation}
relating the ionisation parameter, the density of the absorber and the continuum luminosity $L_{\rm ion}$. In this case the estimate of  $L_{\rm ion}$ (in the energy range between 13.6
eV and 13.6 keV) from the best fit model is of the order of  
 $L_{\rm ion}\sim 7\times 10^{43}$ erg\ s$^{-1}$. Assuming that the thickness of the absorber, $\Delta R$=$  N_{\rm H}/n$, is  smaller than the distance $R_{ion}$ ($\Delta R/R_{ion} <$ 1), we can set an upper limit to the distances of each  ionised absorber, using \suzaku\ observations:
\begin{equation}
\begin{split}
R_{ion} <\frac{L_{\rm ion}  }{N_{\rm H} \xi}
\label{ion_abs_eq2}
\end{split}
\end{equation}
These upper limits are 0.026 pc and 2.72 pc   for the highly  and the mildly  ionised absorber respectively. Despite being upper limits, they suggest that  the likely location of these absorbers does not corresponds to the same radius with respect to the central source. The distance of the first ionised absorber is consistent in being a wind launched from a region located within the Broad Line Region.\\ The higher distance inferred for the mildly ionised absorber is  due to the lower ionisation parameter, with respect to the higly ionised absorber. Its location could correspond to the region of the molecular torus, at a parsec-scale distance from the central source. However as this is an upper limit, we should not exclude the possibility that we are observing across this wind, thus  we are not viewing directly the inner radius, which could be located at sub-parsec scales.\\
Another method to put a constraint on the possible location of these absorbers is to estimate a lower limit on the radial distance by determining the escape radius at which the material will be able to leave the system. This can be determined once we have an estimate of the outflow velocity. Assuming spherical geometry, this radius is:
\begin{equation}
R_{esc}\geq \frac{2GM}{v_{out}^2} \simeq \frac{2c^2R_g}{v_{out}^2}
\end{equation}
where $R_{g}$ is the gravitational radius ($R_g$=GM/$c^2$). Since the measured outflow velocity is $\sim$0.055 c, we obtain that $R_{esc}\geq 660 R_g$. From the literature we have estimates of the mass of the central black hole, ranging from $M_{BH}\sim 1.6 \times 10^7 M_{\bigodot}$ \citep{Woo2002} to $M_{BH}\sim 7.5 \times 10^7 M_{\bigodot}$ \citep{Nikolajuk2004}, so we can infer that $R_g \sim  2.5 - 11.1 \times 10^{12} \rm cm$, thus $R_{esc}\geq 1.6 - 7.3 \times 10^{15} \rm cm \sim 0.0005 - 0.002  $ pc. This means that the wind may have been launched at least from a distance of the order of $10^{-4} - 10^{-3} $pc from the central black hole in order for it to escape, so an origin in the accretion disk or the Broad Line Region is plausible. In particular, this estimate suggests  that the range of location of the first highly ionised absorber is between  $5 \times 10^{-4}$ pc and  0.026 pc in the  \suzaku\ observation.  However, another possibility could be that the wind is part of some aborted outflow,  i.e. a wind with outflow velocity   lower than the escape velocity and thus unable to leave the system; in that case we would only have an estimate of the maximum distance of the ionised absorber but not the minimum distance from which it was launched. \\

This analysis highlights the complexity and possible stratifications of the absorbers intercepting our line of sight. Indeed the observed  spectrum could be explained with physically different scenarios: a unique and multi-phase absorber, where higher density (and lower ionisation) clouds are confined by lower density (higher ionisation) clouds, implying that the location of the two absorbers are actually the same; or a configuration where there are effectively two winds at different physical states and distances, intercepting the line of sight. Neither of these two explanations can be ruled out at the moment, since higher spectral resolution observations are needed.\\

We conclude that the comparison between \suzaku\ and \xmm\ observations of \sorg\ does not show extreme variability, such as the transition from a Compton-thick to a Compton-thin state characterizing ``changing look'' AGN. However we cannot exclude that, given the long time elapsed between observations, we were not able to observe the source during an obscured Compton-thick phase. We do not observe a variation of the primary continuum, despite past RXTE observations which showed brightness variations on time scales down to 1 day  (\citealt{Smith2001}). The satellite ASCA also observed the source (13 years earlier) in a state with flux 7 times lower than the flux observed with \suzaku.  \\
\sorg\ can be placed among the large number of AGN where a non uniform distribution of the circumnuclear absorbing matter determines \nhsym variations in different epochs. This is consistent with recent theoretical models \citep{Nenkova2008a, Nenkova2008b} and works \citep{Elitzur2008,Elitzur2012} that indicate the possible clumpy nature of the torus,  suggesting that  the unified model of AGN is a too simplified scheme. The unified scheme is based on the assumption of a uniform torus, with the same opening angle for all AGN, implying that the viewing angle is the unique factor determining the classification into Type 1 and Type 2 AGN. This is clearly in conflict with  both  the \nhsym\ variability and  infrared observations \citep{Lutz2004,Horst2006} showing that there is no significant difference	in the mid-IR emission, normalized to the X-ray flux, of Type 1 and Type 2 AGN, contrary to the expectation of strong anisotropy of the unified model. As suggested by \cite{Elitzur2012} these observations are compatible with a ''soft-edged'' clumpy torus.

\section{Conclusions}
\label{conclusions}
We presented   the analysis of the X-ray spectrum of \sorg\ obtained by \suzaku\ and compared it to the spectrum observed by \xmm.
\begin{itemize}

\item The best-fit model representing the observed X-ray emission is composed of a primary continuum intercepting three absorbers with different densities and ionisations (including one neutral absorber). We suggest that the location of the neutral absorber and of the mildly ionised absorber are at a parsec scale distance, thus consistent with the location of the putative torus. Instead, the highly ionised absorber appears to be located within $\sim 0.03$ pc from the central source, likely in the Broad Line Region.
\item The  comparison with the \xmm\ observation leads to the conclusions that: 1) the normalization and photon index of the primary and scattered power law do not vary between the two observations, despite such variations were observed in past observations (\citealt{Smith2001}, \citealt{ Awaki2000}); 2) the observed spectral variation requires a change in \nhsym\ of the neutral ($\Delta \rm N_{H} \sim 5.5 \times 10^{22}$\nh) and one of the  ionised absorbers ($\Delta \rm N_{H}  \sim 6.7 \times 10^{22}$\nh).
\item We find the presence of the \feka\ emission line, with equivalent width $EW\sim$81 eV during the \suzaku\ observation, in agreement with a fairly weak   reflection contribution.  During the \xmm\ observation the equivalent width is lower ($EW\sim$ 47 eV), and it could be responding to a past weaker continum, as was observed during earlier \emph{ASCA} observations (\citealt{Awaki2000}). We do not observe any broadening of the \feka\ line as expected for lines produced in the inner regions of the accretion disk, in agreement with the results of \cite{Smith2001} and \cite{Netzer1998}.
\item We detect a weak absorption line with an energy centroid at  $E\sim$7.4 keV, consistent with a possible blueshifted 1s$\rightarrow$2p transition of  Fe\textsc{xxvi} (at 6.95 keV). We infer that the ionised absorber responsible for this feature has an observed outflow velocity of $v_{out}\sim$0.05 c. Higher resolution observations are needed in order to improve the significance of this detection.
\item The long time elapsed  between the two observations does not allow us to infer the time scale of the variability, so we can only determine   upper limits on the distances of the ionised absorbers. However, as the mass of the central black hole is known from previous works, we can also estimate the  minimum escape radius of the outflowing material. The ionisation parameters inferred for the two ionised absorbers suggest that one of  them must be located at sub-parsec scales, and thus in the region ranging from the accretion disk to the BLR, while the second absorber is likely located at a parsec-scale distance.

\end{itemize}

  \section*{ACKNOWLEDGMENTS} 
  This research has made use of the NASA/IPAC Extragalactic Database (NED) which is operated by the Jet Propulsion
Laboratory, California Institute of Technology, under contract with the National Aeronautics and Space
Administration.


\begin{thebibliography}{99}
\bibitem[\protect\citeauthoryear{Ant{\'o}n et al.}{2002}]{Anton2002} Ant{\'o}n S., Thean A., Browne I., Pedlar 
A., 2002, ASPC, 284, 289 
\bibitem[\protect\citeauthoryear{Antonucci}{1993}]{Antonucci} Antonucci, R.\ 1993, \araa, 31, 473 
\bibitem[\protect\citeauthoryear{Akylas et al.}{2002}]{Akylas2002} Akylas, A., Georgantopoulos, I., Griffiths, R. G., Papadakis, I. E., Mastichiadis, A.,
Warwick, R. S., Nandra, K., \& Smith, D. A. 2002, MNRAS, 332, L23
\bibitem[Arnaud(1996)]{xspecref} Arnaud, K.~A.\ 1996, Astronomical Data Analysis Software and Systems V, 101, 17 
\bibitem[\protect\citeauthoryear{Awaki et al.}{2000}]{Awaki2000} Awaki H., Ueno S., Taniguchi Y., Weaver K.~A., 2000, ApJ, 542, 175 


\bibitem[\protect\citeauthoryear{Baird}{1981}]{b1} Baird S.R., 1981,
ApJ, 245, 208
\bibitem[\protect\citeauthoryear{Behar et al.}{2010}]{Behar} Behar, E., Kaspi, S., Reeves, J., et al.\ 2010, \apj, 712, 26 
\bibitem[\protect\citeauthoryear{Bianchi et al.}{2005}]{Bianchi05}Bianchi, S., Guainazzi, M., Matt, G., et al. 2005, A\&A, 442, 185 
\bibitem[\protect\citeauthoryear{Bianchi et al.}{2006}]{Bianchi06} Bianchi, S., Guainazzi, M., \& Chiaberge, M.\ 2006, \aap, 448, 499 
\bibitem[\protect\citeauthoryear{Bianchi et al.}{2009}]{Bianchi09} Bianchi, S., Piconcelli, E., Chiaberge, M., et al.\ 2009, \apj, 695, 781 
\bibitem[\protect\citeauthoryear{Bianchi et al.}{2012}]{Bianchi2012} Bianchi, S., Maiolino,  R., \& Risaliti, G.\ 2012, Advances in Astronomy, 2012,  
\bibitem[\protect\citeauthoryear{Boldt}{1987}]{Boldt} Boldt, E.\ 1987, \physrep, 146, 215
\bibitem[\protect\citeauthoryear{Braito et al.}{2013}]{Braito2013} 
Braito V., Ballo L., Reeves J.~N., Risaliti G., Ptak A., Turner T.~J., 
2013, MNRAS, 428, 2516 
 \bibitem[Dadina(2008)]{Dadina08}Dadina, M. 2008, A\&A, 485, 417

\bibitem[\protect\citeauthoryear{de Vaucouleurs et al.}{1991}]{Vaucouleurs}de Vaucouleurs, G., de Vaucouleurs, A., Corwin, H., Buta, R. J., Paturel, G., \& Fouqu, P. 1991, Third Reference Catalogue of Bright Galaxies (Berlin: Springer)
\bibitem[\protect\citeauthoryear{Done}{2010}]{Done2010}Done C., 2010, arXiv, arXiv:1008.2287
\bibitem[\protect\citeauthoryear{Elitzur}{2008}]{Elitzur2008} Elitzur M., 2008, NewAR, 52, 274 
\bibitem[\protect\citeauthoryear{Elitzur}{2012}]{Elitzur2012} Elitzur M., 2012, ApJ, 747, L33 
\bibitem[\protect\citeauthoryear{Elvis et al.}{2004}]{Elvis04}Elvis, M., Risaliti, G., Nicastro, F., Miller, J. M., Fiore, F., \& Puccetti, S. 2004, ApJ, 615, L25
\bibitem[\protect\citeauthoryear{Elvis}{2012}]{Elvis2012} Elvis 
M., 2012, JPhCS, 372, 012032 
\bibitem[\protect\citeauthoryear{Fukazawa et al.}{2009}]{Fukuzawa2009} Fukazawa Y., et al., 2009, PASJ, 61, 17 

\bibitem[\protect\citeauthoryear{Ghisellini, Haardt, \& Matt}{1994}]{Ghisellini94} Ghisellini G., Haardt F., Matt G., 1994, MNRAS, 267, 743 
\bibitem[\protect\citeauthoryear{Ghisellini et al.}{2004}]{Ghisellini2004} Ghisellini, G.,  Haardt, F., \& Matt, G.\ 2004, \aap, 413, 535 
\bibitem[\protect\citeauthoryear{Gofford et al.}{2013}]{Gofford2013} Gofford J., Reeves J.~N., Tombesi F., Braito V., Turner T.~J., Miller L., Cappi M., 2013, MNRAS, 430, 60 
\bibitem[\protect\citeauthoryear{Gruber et al.}{1999}]{Gruber} Gruber, D.~E., Matteson, J.~L., Peterson, L.~E., \& Jung, G.~V.\ 1999, \apj, 520, 124 

\bibitem[\protect\citeauthoryear{Guainazzi et al.}{2011}]{Guainazzi2011} Guainazzi M., Bianchi S., de La Calle P{\'e}rez I., Dov{\v c}iak M., Longinotti A.~L., 2011, A\&A, 531, A131 





\bibitem[\protect\citeauthoryear{Horst et al.}{2006}]{Horst2006} Horst H., Smette A., Gandhi P., Duschl W.~J., 2006, A\&A, 457, L17 
\bibitem[\protect\citeauthoryear{Jansen et al.}{2001}]{Jansen01}Jansen, F., et al. 2001, A\&A, 365, L1 
\bibitem[\protect\citeauthoryear{Kaastra \& Mewe}{1993}]{Kaastra93}Kaastra, J. S. \& Mewe, R. 1993, A\&AS, 97, 443 
\bibitem[\protect\citeauthoryear{Kaastra et al.}{2000}]{Kaastra2000}Kaastra, J. S., Mewe, R., Liedahl, D. A., Komossa, S., and
Brinkman, A. C. 2000, A\&A 354, L83
\bibitem[\protect\citeauthoryear{Kalberla et 
al.}{2005}]{Kalberla2005} Kalberla P.~M.~W., Burton W.~B., Hartmann D., Arnal E.~M., Bajaja E., Morras R., P{\"o}ppel W.~G.~L., 2005, A\&A, 440, 775 
\bibitem[\protect\citeauthoryear{Kallman et al.}{2004}]{xstar} Kallman, T.~R., Palmeri, P., Bautista, M.~A., Mendoza, C., \& Krolik, J.~H.\ 2004, \apjs, 155, 675 
\bibitem[\protect\citeauthoryear{Kaspi et al.}{2000a}]{Kaspi2000a} Kaspi, S., Brandt, W. N., Netzer, H.et al. 2000a, ApJL 535, L17

 

\bibitem[\protect\citeauthoryear{Khachikian \& Weedman}{1974}]{Khachikian74}Khachikian, E. Y., \& Weedman, D. W. 1974, ApJ, 192, 581
\bibitem[\protect\citeauthoryear{Kokubun et al.}{2007}]{kokubun}Kokubun, M., et al.\ 2007, \pasj, 59, 53 
\bibitem[\protect\citeauthoryear{Koyama et al.}{2007}]{Koyama07} Koyama, K., et al.\ 2007, \pasj, 59, 23 
\bibitem[Kriss et al.(1980)]{Kriss} Kriss, G.~A., Canizares, 
C.~R., \& Ricker, G.~R.\ 1980, \apj, 242, 492 
\bibitem[\protect\citeauthoryear{Laor}{1991}]{laor91} Laor A., 1991, ApJ, 376, 90 
\bibitem[\protect\citeauthoryear{Lutz et al.}{2004}]{Lutz2004} Lutz D., Maiolino R., Spoon H.~W.~W., Moorwood A.~F.~M., 2004, A\&A, 418, 465 


\bibitem[\protect\citeauthoryear{Maiolino et  al.}{2010}]{Maiolino2010} Maiolino, R., Risaliti, G., Salvati, M., et al.\ 2010, \aap, 517, A47 
%
%
\bibitem[\protect\citeauthoryear{Magdziarz \& Zdziarski}{1995}]{pexrav} Magdziarz, P., \& Zdziarski, A.~A.\ 1995, \mnras, 273, 837 
\bibitem[\protect\citeauthoryear{Marchese et 
al.}{2012}]{Marchese2012} Marchese E., Braito V., Della Ceca R., 
Caccianiga A., Severgnini P., 2012, MNRAS, 421, 1803 
\bibitem[\protect\citeauthoryear{Markowitz, Krumpe, \& Nikutta}{2013}]{Markowitz2013} Markowitz A., Krumpe M., Nikutta R., 2013, HEAD, 13, \#108.10 
\bibitem[\protect\citeauthoryear{Matt, Brandt, \& Fabian}{1996}]{Matt96} Matt G., Brandt W.~N., Fabian A.~C., 1996, MNRAS, 280, 823 

\bibitem[Mewe et al.(1985)]{Mewe85} Mewe, R., Gronenschild, E.~H.~B.~M., \& van den Oord, G.~H.~J.\ 1985, \aaps, 62, 197 
\bibitem[\protect\citeauthoryear{Miller  \& Goodrich}{1990}]{Miller1990}  Miller, J. S. \& Goodrich, R. W. 1990, ApJ, 355, 456
\bibitem[\protect\citeauthoryear{Mitsuda et al.}{2007}]{Mitsuda07} Mitsuda, K., et al.\ 2007, \pasj, 59, 1 
\bibitem[\protect\citeauthoryear{Murphy 
\& Yaqoob}{2009}]{Murphy2009} Murphy K.~D., Yaqoob T., 2009, MNRAS, 397, 1549 
\bibitem[\protect\citeauthoryear{Nandra et al.}{2007}]{Nandra2007} Nandra K., O'Neill P.~M., George I.~M., Reeves J.~N., 2007, MNRAS, 382, 194 
\bibitem[\protect\citeauthoryear{Netzer, Turner, \& George}{1998}]{Netzer1998} Netzer H., Turner T.~J., George I.~M., 1998, ApJ, 504, 680 


\bibitem[\protect\citeauthoryear{Neff \& de Bruyn}{1983}]{Neff1983} Neff S.~G., de Bruyn A.~G., 1984, IAUS, 110, 263 
\bibitem[\protect\citeauthoryear{Nenkova et al.}{2008a}]{Nenkova2008a} Nenkova M., Sirocky M.~M., Nikutta R., 
Ivezi{\'c} {\v Z}., Elitzur M., 2008, ApJ, 685, 160 
\bibitem[\protect\citeauthoryear{Nenkova et al.}{2008b}]{Nenkova2008b} Nenkova M., Sirocky M.~M., Ivezi{\'c} {\v 
Z}., Elitzur M., 2008, ApJ, 685, 147 
\bibitem[\protect\citeauthoryear{Nikolajuk, Papadakis, \& Czerny}{2004}]{Nikolajuk2004} Nikolajuk M., Papadakis I.~E., Czerny B., 2004, MNRAS, 350, L26 
\bibitem[\protect\citeauthoryear{Palmeri et al.}{2003}]{Palmeri} Palmeri, P., Mendoza, C., Kallman, T.~R., Bautista, M.~A., \& Mel{\'e}ndez, M.\ 2003, \aap, 410, 359 


\bibitem[\protect\citeauthoryear{Puccetti et al.}{2007}]{Puccetti} Puccetti, S., Fiore,  F., Risaliti, G., et al.\ 2007, \mnras, 377, 607 
\bibitem[\protect\citeauthoryear{Risaliti}{2000}]{Risa2002} Risaliti, G.\ 2002, \aap, 386, 379 
\bibitem[\protect\citeauthoryear{Risaliti et al.}{2002}]{Risaliti2002} Risaliti, G., Elvis, M., \& Nicastro, F.\ 2002, \apj, 571, 234 
\bibitem[\protect\citeauthoryear{Risaliti et al.}{2005}]{Risaliti05} Risaliti, G., Bianchi, S., Matt, G., Baldi, A., Elvis, M., Fabbiano,G., \& Zezas, A.\ 2005,  \apjl, 630, L129 
\bibitem[\protect\citeauthoryear{Risaliti et al.}{2007}]{Risaliti07} Risaliti, G., Elvis,  M., Fabbiano, G., et al.\ 2007, \apjl, 659, L111 
\bibitem[\protect\citeauthoryear{Risaliti et al.}{2009}]{Risa2009a} Risaliti, G., et al.\ 2009, \mnras, 393, L1 
\bibitem[\protect\citeauthoryear{Risaliti et al.}{2010}]{Risaliti10} Risaliti, G., Elvis,  M., Bianchi, S., \& Matt, G.\ 2010, \mnras, 406, L20 
\bibitem[\protect\citeauthoryear{Risaliti}{2010}]{Risaliti2010a}Risaliti, G. 2010, in American Institute of Physics Conference Series, Vol. 1248, American Institute of Physics Conference Series, ed. A. Comastri, L. Angelini, \& M. Cappi, 351?354

\bibitem[\protect\citeauthoryear{Rivers, Markowitz, 
\& Rothschild}{2011}]{Rivers11} Rivers E., Markowitz A., Rothschild R., 2011, ApJ, 742, L29 
\bibitem[\protect\citeauthoryear{Rivers, Markowitz,  \& Rothschild}{2013}]{Rivers13} Rivers E., Markowitz A., Rothschild R., 2013, ApJ, 772, 114 



\bibitem[\protect\citeauthoryear{Simpson et al.}{1996}]{Simpson1996} Simpson C., Mulchaey J.~S., Wilson A.~S., 
Ward M.~J., Alonso-Herrero A., 1996, ApJ, 457, L19 
\bibitem[\protect\citeauthoryear{Singh et al.}{2011}]{Singh2011}Singh V., Shastri P., Risaliti G., 2011, A\&A, 532, A84 
\bibitem[\protect\citeauthoryear{Smith et al.} {2001}]{Smith2001}Smith, D. A., Georgantopoulos , I., Warwick, R. S., 2001, ApJ,
550, 635


\bibitem[\protect\citeauthoryear{Str\"{u}der et al.}{2001}]{Struder01}Str\"{u}der, L., et al. 2001, A\&A, 365, L5
\bibitem[\protect\citeauthoryear{Takahashi et al.}{2007}]{Takahashi07}Takahashi, T., et  al.\ 2007, \pasj, 59, 35 
%
\bibitem[\protect\citeauthoryear{Tombesi et al.}{2010}]{Tombesi2010b}Tombesi F., Cappi M., Reeves J. N., Palumbo G. G. C., Yaqoob T., Braito
V., Dadina M., 2010, A\&A, 521, 57
\bibitem[\protect\citeauthoryear{Tombesi et al.}{2011}]{Tombesi2011}Tombesi F., Cappi M., Reeves J. N., Palumbo G. G. C.,
Braito V., Dadina M., 2011, ApJ accepted
\bibitem[\protect\citeauthoryear{Tombesi et 
al.}{2012}]{Tombesi12} Tombesi F., Cappi M., Reeves J.~N., Nemmen 
R.~S., Braito V., Gaspari M., Reynolds C.~S., 2012, arXiv, arXiv:1212.4851 
\bibitem[\protect\citeauthoryear{Tombesi et al.}{2013}]{Tombesi2013} Tombesi F., Cappi M., Reeves J.~N., Nemmen R.~S., Braito V., Gaspari M., Reynolds C.~S., 2013, MNRAS, 430, 1102 
\bibitem[\protect\citeauthoryear{Turner et al.}{2000}]{Turner2000} Turner T.~J., Perola G.~C., Fiore F., Matt G., George I.~M., Piro L., 
Bassani L., 2000, ApJ, 531, 245 




%
\bibitem[\protect\citeauthoryear{Turner et al.}{ 2001}]{Turner01}Turner, M., et al. 2001, A\&A, 365, L27
\bibitem[\protect\citeauthoryear{Turner \& Miller}{2009}]{Turner09}Turner T.J., Miller L., 2009, A\&ARv, 17, 47
5\bibitem[\protect\citeauthoryear{Turner et al.}{2009}]{Turner09} Turner, T.~J., Miller, 
L., Kraemer, S.~B., Reeves, J.~N., \& Pounds, K.~A.\ 2009, \apj, 698, 99 
\bibitem[\protect\citeauthoryear{Turner et al.}{2012}]{Turner2012} Turner, T.~J., Miller,  L.,  \& Tatum, M.\ 2012, American Institute of Physics Conference Series, 1427, 165 
\bibitem[\protect\citeauthoryear{Unger et al.}{1984}]{Unger1984} Unger S.~W., Pedlar A., Neff S.~G., de Bruyn A.~G., 1984, MNRAS, 209, 15P 

\bibitem[\protect\citeauthoryear{Urry \& Padovani}{1995}]{Urry1995} Urry, C.~M., \& Padovani, P.\ 1995, \pasp, 107, 803

\bibitem[\protect\citeauthoryear{Warwick et al.}{1989}]{Warwick89}Warwick, R. S., Koyama, K., Inoue, H., Takano, S., Awaki, H., \& Hoshi, R. 1989, PASJ, 41, 739
\bibitem[\protect\citeauthoryear{Wilms, Allen, \& McCray}{2000}]{Wilms2000} Wilms J., Allen A., McCray R., 2000, ApJ, 542, 914 

\bibitem[\protect\citeauthoryear{Woo \& Urry}{2002}]{Woo2002} Woo J.-H., Urry C.~M., 2002, ApJ, 579, 530 
\bibitem[\protect\citeauthoryear{Xue et al.}{1998}]{Xue98} Xue S.-J., Otani C., Mihara T., Cappi M., Matsuoka M., 1998, PASJ, 50, 519 

%
%
%
%
\end{thebibliography}
 \end{document}